\documentclass[fleqn,usenatbib]{mnras}
\usepackage{float}
\usepackage{comment}
\usepackage{xcolor}
\usepackage{soul}
\usepackage{amsmath}
\usepackage{hyperref}
\usepackage{amssymb}
\usepackage{graphicx}
\hypersetup{linkcolor=blue,citecolor=blue,filecolor=cyan,urlcolor=magenta}

\renewcommand{\ion}[2]{\textnormal{#1\,\textsc{\romannumeral #2}}}
\newcommand{\msun}{M$_\odot$ }
\newcommand{\psqcm}{cm$^{-2}$ }

\newcommand{\hi}{\ion{H}{1} }

\newcommand{\civ}{\ion{C}{4} }

\newcommand{\ovi}{\ion{O}{6} }
\newcommand{\ovii}{\ion{O}{7} }
\newcommand{\oviii}{\ion{O}{8} }
\newcommand{\oix}{\ion{O}{9} }

\setstcolor{cyan}

\setstcolor{red}

\defcitealias{Birnboim2003}{BD03}
\defcitealias{Bromberg_2025}{B+25}
\defcitealias{Sarkar2021a}{S+21}
\newcommand{\hydra}{\textsc{hydra }}

\author[Bromberg et al.]{
Itai Bromberg,$^{1}$\thanks{E-mail: itai.bromberg@mail.huji.ac.il}
Kartick C. Sarkar,$^{2}$
Orly Gnat,$^{1}$
Yuval Birnboim$^{1}$
\\
$^{1}$Centre for Astrophysics and Planetary Science, Racah Institute of Physics, The Hebrew University, Jerusalem 91904, Israel\\
$^{2}$Raman Research Institute, Bangalore, India
}

\title[Non-Equilibrium Ionisation and Stability in Virial Shocks]{Non-Equilibrium Ionisation in Photoionised Haloes: Implications for Shock Stability and Absorption-Line Signatures}

\begin{document}

\maketitle

\begin{abstract}

%%%%%%%%%%%%%%%%%%%%%%%%%%%%%%%%%%%%%%%%%%%%%%%%%%%%%%%%%%%%%%%%%%%%%%
%%%%%%%%%%%%%%%   A   B   S   T   R   A   C   T   %%%%%%%%%%%%%%%%%%%%
%%%%%%%%%%%%%%%%%%%%%%%%%%%%%%%%%%%%%%%%%%%%%%%%%%%%%%%%%%%%%%%%%%%%%%
We investigate the impact of non-equilibrium ionisation (NEI) and the metagalactic radiation-field on the thermal evolution, virial shock stability, and absorption signatures of gas surrounding galaxies. Using 1D, spherically symmetric hydrodynamical simulations with an extended version of the \hydra code, we follow dark-matter growth, gas dynamics, time-dependent ionisation and cooling in the presence of the UV background. We explicitly track all ions of H, He, C, N, O, Ne, Mg, Si, S, and Fe in haloes of mass $10^{11}$–$10^{13}$~M$_\odot$ from $z=100$ to $z=0$.
Without a UV background, NEI enhances post-shock cooling due to underionised gas, reducing pressure support and raising the minimum mass for stable shock formation. 
Including the UV background pre-ionises the IGM, suppressing NEI, and restoring the CIE threshold.
We find that the IGM temperatures deviate from thermal equilibrium due to adiabatic expansion and collapse, while ionisation remains close to equilibrium in the presence of a UV background, except in transient rapidly cooling regions where NEI occurs.
We compute absorption columns of \ion{O}{6}, \ion{C}{4}, and \ion{H}{1}, showing that a photoionised IGM may produce substantial warm-ion columns extending beyond $R_{\rm vir}$, including \ion{O}{6} column densities comparable to observed values. Our models indicate weak halo-mass dependence and extended distributions.
We also find that $z\gtrsim3$ haloes can produce \ion{C}{4} ($N_{\rm C,IV}\sim10^{13-15}$~cm$^{-2}$) and \ion{H}{1} ($N_{\rm H,I}\sim10^{15-17}$~cm$^{-2}$) columns out to $\sim10R_{\rm vir}$.
Our results highlight the role of the UV background in regulating the thermal state and observable signatures of the gas surrounding galaxies, and emphasize the importance of accounting for IGM contributions when interpreting CGM absorption-line observations.

\end{abstract}

% %%%%%%%%%%%%%%%%%%%%%% MAIN BODY  %%%%%%%%%%%%%%
\section{Introduction} %\label{sec:intro}

In the present-day universe, a large fraction of the baryonic matter resides outside galaxies, in the diffuse gas that fills the regions between and around them. This gas forms the circumgalactic medium (CGM) and the intergalactic medium (IGM). The CGM plays a central role in galaxy evolution, regulating both the accretion of gas onto galaxies and the ejection of material through outflows \citep{Prochaska2011, Tumlinson2017}.

Early theoretical work \citep{ro77,wr78,silk77} 
established that hot gaseous haloes  can naturally form as intergalactic gas is shock-heated while accreting into the gravitational potential wells of forming galaxies. The ensuing thermal and chemical evolution of this gas is shaped by a complex interplay of hydrodynamics, radiative cooling, and ionisation processes. As a result, the CGM, confined within the virial shock radius, becomes a multiphase and dynamically rich environment \citep[e.g.][]{Shull_2003, Werk_2016, Oppenheimer2018}.

Observationally, the distribution of ionised gas around galaxies has been extensively probed through both UV absorption-line studies and X-ray observations. Early work with the \textit{COS}-Halos survey \citep{Tumlinson2011} revealed widespread \ovi\ absorption in the CGM, with relatively flat column density profiles out to $\sim R_{\rm vir}$. Subsequent studies, including IMACS \citep{Johnson_2015} and CGM$^2$ \citep{Tchernyshyov_2022}, have extended these measurements and suggest that \ovi\ absorption may persist to large impact parameters, in some cases well beyond the virial radius. Additional observations 
\citep{Burchett+19_COS_NeVIII, Ng+19_OVI_CGM_obs, Zheng+19_MW_CGM_COS} further indicate that the CGM is extended and multiphase, with contributions from both warm and hot gas. Complementary X-ray studies \citep{gupta12, yao12, nicastro13} provide evidence for a pervasive hot halo component surrounding galaxies. However, the radial extent and physical origin of this gas remain uncertain: while some studies suggest that highly ionised material is largely confined within the virial radius \citep[e.g.][]{Johnson_2015, Qu2024_CUBS}, others find evidence for substantial ionised gas extending well into the surrounding IGM \citep{Prochaska2011, Tchernyshyov_2022}.

Our theoretical understanding of the CGM is built upon a combination of complementary modeling approaches that span a wide range of physical scales and levels of complexity. Cosmological simulations provide a global framework for studying galaxy formation and the CGM in its cosmological context, capturing structure growth and environmental diversity, but necessarily rely on subgrid prescriptions to model unresolved processes such as star formation, feedback, and microphysical effects \citep[e.g.][]{Nelson2018,Oppenheimer2018,Ho2021,Appelby2022,Strawn_2024}. At the other extreme, 
low-dimensional models and simulations focus on specific physical processes, allowing for efficient exploration of parameter space and clearer interpretation of how individual mechanisms—such as cooling, photoionisation, or feedback—shape the observable properties of the CGM \citep[e.g.][]{Faerman_2017,Faerman_2020,Stern2016, Stern2019,stern18,McQuinn_2018,Qu_2018b}. 

Warm-ion absorption around galaxies is traditionally attributed to hot, collisionally ionised gas within the CGM. In our previous work \cite[][hereafter B+25]{Bromberg_2025}, we augmented that interpretation, showing that a significant fraction of these detections may instead originate in cooler, photoionised gas in the surrounding IGM, particularly for lower-mass haloes.
In that work, we used a spherically symmetric analytic toy-model
of halo formation to demonstrate that warm-ion absorption arises naturally in the IGM as it responds to the UV background radiation field. 

The \citetalias{Bromberg_2025} model, which assumed ionisation equilibrium, showed that photoionised IGM alone could account for a significant fraction of the observed \ovi columns. This contribution arises from gravitational focusing, which enhances the gas density in the IGM outside of the shock radius, and allows warm ions to form under photoionisation equilibrium (PIE) conditions. These findings emphasised the importance of including the contribution of IGM when interpreting CGM absorption-line observations. 
However, the treatment in \citetalias{Bromberg_2025} was highly idealized, relying on a static CGM model and neglecting several key physical processes, including full hydrodynamical evolution and departures from ionisation equilibrium. Both effects are expected to influence the thermal history, ionisation structure, and resulting observable signatures of the gas.

Using one-dimensional spherical hydrodynamic and N-body simulations \citet[hereafter BD03]{Birnboim2003} showed that haloes must satisfy a stability criterion ($M_{\rm vir} \gtrsim 10^{12}$~M$_\odot$) for a virial shock to form around them. In their model, radiative cooling was computed using collisional ionisation equilibrium (CIE) tables from \citet{Sutherland1993}. They found that halo masses below a critical mass $\approx 3\times10^{11}$~M$_\odot$ \citep[with the exact threshold depending on metallicity and redshift; see ][]{Dekel2006}, a stable shock cannot be maintained. In these systems, rapid cooling behind the shock reduces the post-shock pressure support, causing the shock to collapse or preventing it from forming altogether. 
This behaviour has significant implications for CGM observational signatures, particularly in unstable haloes with masses below the critical mass. In such cases, the shock radius lies well within the virial radius $-$ potentially, as low as the edge of the galaxy $-$ so that a substantial fraction of the gas within $R_{\rm vir}$ remains part of the infalling IGM rather than a virialized hot halo. These results highlight the importance of a consistent hydrodynamic treatment in modelling the structure and evolution of the CGM and its surrounding environment.

In rapidly evolving regions, such as shock fronts and post-shock cooling layers, where heating or cooling proceeds faster than ionisation or recombination, non-equilibrium ionisation (NEI) becomes significant. In these conditions, the ion fractions of key species such as \ovi\ and \civ\ can deviate substantially from their equilibrium values, as ionisation and recombination processes lag behind the thermal evolution. This affects both emission and absorption-line diagnostics. In addition, departures from equilibrium can modify radiative cooling rates, with cooling enhanced in underionised gas and suppressed in overionised gas \citep{Gnat2007, Gnat2009, Sarkar2021a,Sarkar2021b}. Consequently, time-dependent ionisation modelling may be important
for accurately predicting ion abundances and thermal evolution, and for interpreting CGM observations in environments undergoing rapid heating and cooling transitions.

However, if the gas is exposed to an ambient ionising radiation field, such as the metagalactic UV background, the impact of non-equilibrium ionisation may be significantly reduced \citep{Vasiliev2011, Oppenheimer_2013, Gnat2017}. In such cases, although NEI processes remain important for a physically consistent description of the gas, their effect on observable signatures may be limited \citep{Arya2026}.

In this work, we reexamine the findings of B+25 using a more realistic and physically complete model. Instead of the analytic, dark-matter-only toy model, we employ a one-dimensional simulation within a $\Lambda$CDM cosmology, which more accurately captures halo growth and self-consistently follows the gas hydrodynamics, including shock formation.
We build upon the Lagrangian code \hydra introduced in \citetalias{Birnboim2003}, extending it to include additional physical ingredients. First, we incorporate a time-dependent ionisation network that follows the non-equilibrium ion fractions, as well as the associated cooling and heating rates, in the evolving gas. For this purpose, we adopt the ionisation network presented in \citet[][hereafter S+21]{Sarkar2021a}. Second, we include the effects of the redshift-dependent metagalactic UV background, which photoionises and heats the gas.

The enhanced version of \hydra allows us to toggle between models with and without a UV background, and with either equilibrium or non-equilibrium ionisation, enabling a systematic exploration of their impact on shock stability and observable signatures. We simulate haloes across a range of masses, solving the coupled equations of hydrodynamics, radiative cooling and heating, and ionisation evolution. This approach allows us to follow the spatial and temporal evolution of the gas density, temperature, and ionisation state both within and beyond the shock radius.

The goals of this study are threefold: (1) to evaluate the impact of departures from ionisation equilibrium on the stability of virial shocks; (2) to assess the role of the UV background in regulating shock stability; and (3) to evaluate the resulting absorption-line column densities and to quantify the contribution of infalling IGM gas to absorption-line observations. As we use one-dimensional simulations, this work is not intended to model the full multiphase CGM, but rather to isolate how ionisation physics alone affects shock stability and large-scale ion distributions.

The structure of the paper is as follows. In \S~2, we describe our model, including the governing hydrodynamic and ionisation equations, the implementation of the UV background, the numerical methods, and the explored parameter space. In \S~3, we present our results regarding shock stability and examine the roles of non-equilibrium ionisation and the UV background. In \S~4, we analyse the resulting thermal and ionisation structure of the gas in and around haloes. In \S~5, we investigate the absorption-line signatures of our models, focusing on the roles of NEI and UV radiation and demonstrating the significant contribution of the IGM to high-ion column densities. In \S~6, we discuss the main limitations of our model, and in \S~7 we summarize our conclusions.

%%%%%%%%%%%%%%%%%%%%%%%%%%%%%%%%%%%%%%%%%%%%%%%%%%%%%%%%%%%%%%%%%%%%%%%%%%%%%%%%%%%%%%%%%%%%%%%%%%%%%%%%%%%%%%%%%%%%%%%%%%%%%%%%%%%%%%%%%%%%%%%%%%%%%%%%%%%%%%%%%%%%%%%%%%%%%%%%%%%%%%%%%%%%%%%%%%%%%%%%%%%%%%%%%%%%%%%%%%%%%%%%%%%%%%%%%%%%%%%%%%%%%%%%%%%%%%%%%%%%%%%%%%%%%%%%%%%%%%%%%%%%%%%%%%%%%%%%%%%%%%%%%%%%%%%%%%%%%%%%%%%%%%%%%%%%%%%%%%%%%%%%%%%%%%%%%%%%%%%%%%%%%%%%%%%%%%%%%%%%%%%%%%%%%%

\section{Method} \label{sec:method}

We use \hydra (\citetalias{Birnboim2003}, \citealp{Dekel2006,Birnboim2007}), a one-dimensional, spherically symmetric Lagrangian hydrodynamics code, to simulate the evolution of gas and dark matter in spherical haloes. The code follows the radial evolution of concentric shells of gas and collisionless dark matter under gravity, while solving the Euler equations for the gas. We also couple the hydrodynamics to a time-dependent ionisation network \citepalias{Sarkar2021a}, allowing us to follow the non-equilibrium ionisation state of the gas when desired. This enables us to compute the corresponding radiative cooling and photoabsorption heating due to the metagalactic UV background.

In the following subsections, we describe the method used to follow the evolution of the dark matter (\S~\ref{subsec:dark-matter}), gas hydrodynamics (\S~\ref{subsec:hydrodynamics}), ionisation states (\S~\ref{subsec:NEImethod}), and the ionising background radiation (\S~\ref{subsec:UV}). We then present the suite of simulations used to study the stability of, and absorption-line signatures around, spherical haloes (\S~\ref{subsec:suite}).

%%%%%%%%%%%%%%%%%%%%%%%%%%%%%%%%%%%%%%%%%%%%%%%%%%%%%%%%%%%%%%%%%%%%%%%%%%%%%%
\subsection{The Evolution of the Dark Matter} \label{subsec:dark-matter}

As described above, in \hydra, dark matter is represented by collisionless shells and evolves under the influence of gravity. 
The gravitational force includes contributions from both gas and dark matter, with a small central smoothing length introduced to avoid numerical singularities. Angular momentum is included through a centrifugal term in the equation of motion, with a specific angular momentum assigned to each shell and conserved during the evolution. The simulations adopt a $\Lambda$CDM cosmology with $\Omega_m=0.3$, $\Omega_\Lambda=0.7$, $h=0.7$ and $\sigma_8=1$, and include cosmic expansion due to the cosmological constant.
The equation of motion reads,

\begin{equation}
    \frac{d^2r}{dt^2} = -\frac{G M_{\rm DM}(<r) + G M_{\rm gas}(<r)}{(r+a)^2} + \frac{j^2}{R^3} + h^2 \Omega_\Lambda r
\end{equation}
where $M_{\rm DM}(<r)$ and $M_{\rm gas}(<r)$ are the total dark matter and baryonic mass inside radius $r$, and the softening length, $a$ ($= 50$ pc), prevents a singularity at $r=0$.
The term $j$ is the specific angular momentum of the shell, providing centrifugal support.
The last term ($h^2\Omega_\Lambda r$) represents the repulsive acceleration due to the cosmological constant, where $h$ is the dimensionless Hubble parameter, and $\Omega_\Lambda$ is the cosmological constant density parameter. The equation of motions of the dark matter, as well as of the baryons (\S\ref{subsec:hydrodynamics}), are solved by using a fourth order Runge–Kutta scheme. If the higher order corrections exceed some threshold fraction over the second-order correction, the simulation reverts to the beginning of the time-step and proceeds with a smaller time-step to ensure convergence in time-steps.

As in \citetalias{Birnboim2003}; \cite{Dekel2006,Birnboim2007}, the simulations are initialized at a high redshift ($z=100$) with a small spherically symmetric density perturbation embedded in an expanding cosmological background, which is reflected in the initial velocity profile. 
The initial gas and dark matter distributions follow the same perturbation profile, which is taken to be proportional to the linear two-point correlation function of the assumed cosmology, representing a typical fluctuation in a Gaussian random field. The amplitude of the perturbation determines the collapse time of different mass shells. 

%%%%%%%%%%%%%%%%%%%%%%%%%%%%%%%%%%%%%%%%%%%%%%%%%%%%%%%%%%%%%%%%%%%%%%%%%%%%%%%%%%%%%%%%%
%%%%%%%%%%%%%%%%%%%%%%%%%%%%%%%%%%%%%%%%%%%%%%%%%%%%%%%%%%%%%%%%%%%%%%%%%%%%%%%%%%%%%%%%%
\subsection{Evolution of the Gas} \label{subsec:hydrodynamics}

For the gas, \hydra follows the hydrodynamic properties of the baryonic shells.
It solves the following set of one dimensional Lagrangian hydrodynamic equations, for each shell,
\begin{align}
&\frac{d^2 r}{dt^2} = -\frac{1}{\rho}\nabla(P+\sigma) +\\
&\quad\quad\quad - \frac{G\left[M_{\rm DM}(<r)+M_{\rm gas}(<r)\right]}{(r+a)^2} + \nonumber \\
&\quad\quad\quad + \frac{j^2}{r^3} + h^2 \Omega_\Lambda r \nonumber\\
&\rho = \frac{dM_{\rm gas}}{4\pi r^2\,dr} \\
&\frac{de}{dt} = \frac{P+\sigma}{\rho^2}\frac{d\rho}{dt} - q \\
&P = (\gamma-1)\rho e \, .
\end{align}
where, $P$, $e$, $\rho$, $q$ and $\gamma$ denote the gas pressure, specific internal energy, gas density, net radiative cooling rate per unit mass, and the adiabatic index, respectively. Here, $\sigma$ is the artificial viscosity used to capture shocks \citepalias{Birnboim2003}. 

The hydrodynamic evolution is computed using an operator-splitting approach. At each time step, the hydrodynamic equations are first advanced without the radiative source term 
$q$, updating the gas density, velocity, and internal energy due to the pressure and gravity forces. The thermodynamic state of the gas is then propagated over the full time-step at constant density while the ionisation network is evolved (see \S~\ref{subsec:NEImethod}). This yields the updated internal energy (temperature), ion fractions and the corresponding radiative cooling and photoabsorption heating rate $q$. 

We consider $1000$ baryonic shells, and $10000$ dark matter shells in each simulation to properly resolve the virial shock formation.

\subsubsection{Metallicity Profile} \label{subsec:metallicity}

An important parameter in the calculation of radiative losses is the gas metallicity, since the radiative cooling rates depend strongly on the abundance of heavy elements. This, in turn, affects the evolution of the ionisation network (see \S~\ref{subsec:NEImethod}). \hydra does not self-consistently track the metallicity of the gas. Instead, the metallicity is 
prescribed as a function of position. 
We assume a metallicity profile consisting of an infalling IGM metallicity ($Z_{\rm IGM}$) and a CGM metallicity ($Z_{\rm CGM}$):
\begin{equation}
    Z(r) = \begin{cases} 
        \text{$Z_{\rm CGM}=0.3$} & \text{$,~~~r\leq r_s$} \\
        \text{$Z_{\rm IGM}=0.1$} & \text{$,~~~r>r_s$} \\
    \end{cases}
\end{equation}
 where $r_s$ is the shock radius. 
While the metallicity in and around haloes remains an active area of research, both observations and simulations of the CGM generally find metallicities in the range $0.1$--$1\,Z_\odot$ \citep{Lehner_2019, Wotta_2019, Cook_2025, Grayson_2025, Oppenheimer_2025}, with simulations typically favouring the lower end of this range. Our choice of $Z_{\rm CGM}=0.3\,Z_\odot$, which lies near the middle of this range, therefore provides a representative value for the shocked-halo gas.
Metallicity estimates for gas outside the virial radius, extending out to the turnaround radius, are more limited. However, numerical simulations suggest typical values of $Z\sim 0.1$ Z$_\odot$ in these regions \citep{Grayson_2025,Oppenheimer_2025,Arya2026}.

In our simulations, the metallicity is initially set everywhere to $Z = Z_{\rm IGM}$. To model the enrichment of halo gas, \hydra tracks (see \S\ref{sec:shock-stability})
the shock radius, $r_{\rm s}$, as a function of time and assigns a metallicity $Z_{\rm CGM}$ to a gas shell once it crosses the shock radius. This is a phenomenological prescription, used to separate the IGM and CGM metallicities.
After a Lagrangian shell is shocked, its metallicity is kept fixed at $Z_{\rm CGM}$ for the remainder of the simulation, even if the shock later recedes and the shell temporarily moves to $r > r_{\rm s}$. This treatment is physically motivated, as gas that has been enriched is not expected to lose its metal content if the shock becomes temporarily unstable. In practice, such cases are rare and typically last only a few time steps, with negligible impact on our results.

\subsection{Time-Dependent Ionisation and Cooling} \label{subsec:NEImethod}
For tracking the evolution of the non-equilibrium ionisation, radiative cooling, and photoabsorption heating, we use the ionisation network (IN) presented in \citetalias{Sarkar2021a}. 
Here we use the IN to obtain the time-dependent ionisation fractions of H, He, C, N, O, Ne, Mg, S, Si, and Fe in each baryonic shell (spanning a network of 104 ions).  
The non-equilibrium ion fractions are then used to calculate the associated radiative cooling and photoabsorption heating rates. For the IN, we solve the following set of coupled differential equations,
\begin{equation}\label{eq:ion_fractions}
\frac{dx_{k,i}}{dt}=S_{k,i}(\{x\},Z,z,T)
\end{equation}
where $x_{k,i}$ is the ion fraction of ion $i$ of element $k$, 
and $S_{k,i}$ is the source term for an ion $i$ of element $k$. The source term $S_{k,i}$ takes into account the total radiative and dielectric recombination, collisional ionisation, photoionisation, Auger ionisation, and charge-transfer reactions. Details of the calculation of the source term ($S_{k,i}$) and the solution of this set of equations are described in \citepalias{Sarkar2021a}.

The IN also calculates, for each shell and time, the radiative cooling and photoabsorption heating rates for a given set of (non-equilibrium) ion fractions, metallicity, density, and the radiation field. The radiative cooling $\mathcal{L}$ (erg s$^{-1}$ cm$^{-3}$) is comprised of free-free emission, collisionally excited- and recombination-line emissions, and is given by:
\begin{equation}
    \mathcal{L}=n_e\sum_{k,i}n_{k,i}\Lambda_{k,i}(T)
\end{equation}
where $n_e$ and $n_{k,i}$ are the electron and ion densities, respectively, and $\Lambda_{k,i}$ is the ion radiation efficiency taken from pre-computed tables from CLOUDY-17 \citep{Ferland2017} among other previous works. 
The heating includes the photoabsorption-heating (PH) and charge transfer (CT) terms:
\begin{equation}
    \mathcal{H}=H_{ph}+H_{ct}.
\end{equation}
The calculation of $H_{ph}$ and $H_{ct}$ follows eq. 15 in \citepalias{Sarkar2021a}.

We invoke the IN module at the end of every \hydra hydrodynamic step.  
We evolve the ion fractions and energy (after cooling or heating) using the coupled set of ionisation equations and the energy equation. 
We sub-cycle the hydrodynamic time step to reduce integration errors, and if the equations become stiff the code reverts to analytic matrix inversion for these ions.
Since the IN calculation dominates the computation time, we implemented a parallelized MPI-based solution where at every ionisation time-step each shell is assigned a task for the calculation. On completion, the main ("hydro") thread collects the results and assigns a new shell to that thread until all shells are done. We use between 32 and 64 nodes per calculation, yielding a considerable acceleration in computation time.

Another mode of calculation is the equilibrium model, in which ion fractions are set to their equilibrium values ($d{x}/dt = 0$) at the local temperature and density. The temperature is then evolved using the corresponding equilibrium cooling rate, integrated with subcycled timesteps.

\subsection{UV Background Evolution}\label{subsec:UV}

We include the UV background radiation field as a function of redshift using the tables from \citet[][hereafter, HM12]{Haardt2012}, employing linear interpolation between tabulated redshifts. The spectral range is discretized into $40$ frequency bins, spanning an energy range of $5-2,500$ eV, while maintaining the primary ionisation edges of H, He, C, N, and O. A list of frequencies used in this work is given in \citepalias{Sarkar2021a}. 
Using an isochoric gas-cooling test \citep[cf.][]{Gnat2007}, we verified that this reduced frequency sampling does not noticeably affect the ion fractions or the overall cooling and heating rates.

\subsection{Simulation Suite}\label{subsec:suite}

We compute the coupled hydrodynamic, thermal, and ionisation-state evolution for haloes with present-day ($z=0$) dark-matter masses $\rm M_h = 10^{11},\, 3\times10^{11},\, 10^{12},\, 3\times10^{12}$, and $10^{13}\,M_\odot$. For each halo mass, we consider four different physical scenarios. 
The first scenario is similar to that considered in \citetalias{Birnboim2003} (albeit with the newer cooling rates): the ionisation states are assumed to be in ionisation equilibrium and no UV background radiation is included. 
We refer to this scenario as `Eq'. 
We then consider departures from equilibrium ionisation by following the time-dependent ion fractions and cooling rates, but still without a background radiation field. 
These models are referred to as `nEq'. 
Next, we examine the impact of the UV background radiation on halo evolution, both in the equilibrium case (`Eq+UV') and in the non-equilibrium case (`nEq+UV').
We therefore consider four models: 'Eq', 'nEq', 'Eq+UV', and 'nEq+UV', as summarised in Table~\ref{tab:models}.
For the equilibrium models, the ion fractions depend only on the local density, temperature, and ionising radiation field, rather than being history-dependent as implied by equation~\ref{eq:ion_fractions}.

\begin{table}
\centering
\caption{Summary of the physical scenarios explored in this work.}
\begin{tabular}{lcc}
\hline
Model & UV Background & Ionisation  \\
\hline
Eq        & No  & Equilibrium \\
nEq       & No  & Non-equilibrium \\
Eq+UV     & Yes & Equilibrium \\
nEq+UV    & Yes & Non-equilibrium\\
\hline
\end{tabular}
\label{tab:models}
\end{table}

\section{Results: Shock stability} \label{sec:shock-stability}
A stable shock is a shock that is supported by sufficient thermal pressure. The shock, however, may lose the thermal energy via radiative cooling and may not be able to sustain itself against the gravitational pull of the halo. Such unstable shocks are often noticed in low mass haloes ($M_h \lesssim 3\times10^{11}$ \msun) where the virial velocity is such that the resulting shock temperature is $\lesssim 10^5$ K, and therefore, the shock suffers from extreme radiative cooling. The stability of the shock is discussed in detail in \citetalias{Birnboim2003}. This work is an extension of \citetalias{Birnboim2003}, but includes deviations from ionisation equilibrium, and the redshift-dependent UV background from HM12. 
In \citetalias{Birnboim2003}, assuming equilibrium cooling, an analytic criterion for shock formation was derived as a function of infall velocity, density, and metallicity, which can be mapped to halo mass and redshift in a cosmological context. In contrast, with the inclusion of time-dependent non-equilibrium processes, the criterion for shock formation can no longer be derived analytically and must instead be determined numerically.

\begin{figure*}
\centering
\includegraphics[width=0.98\textwidth]{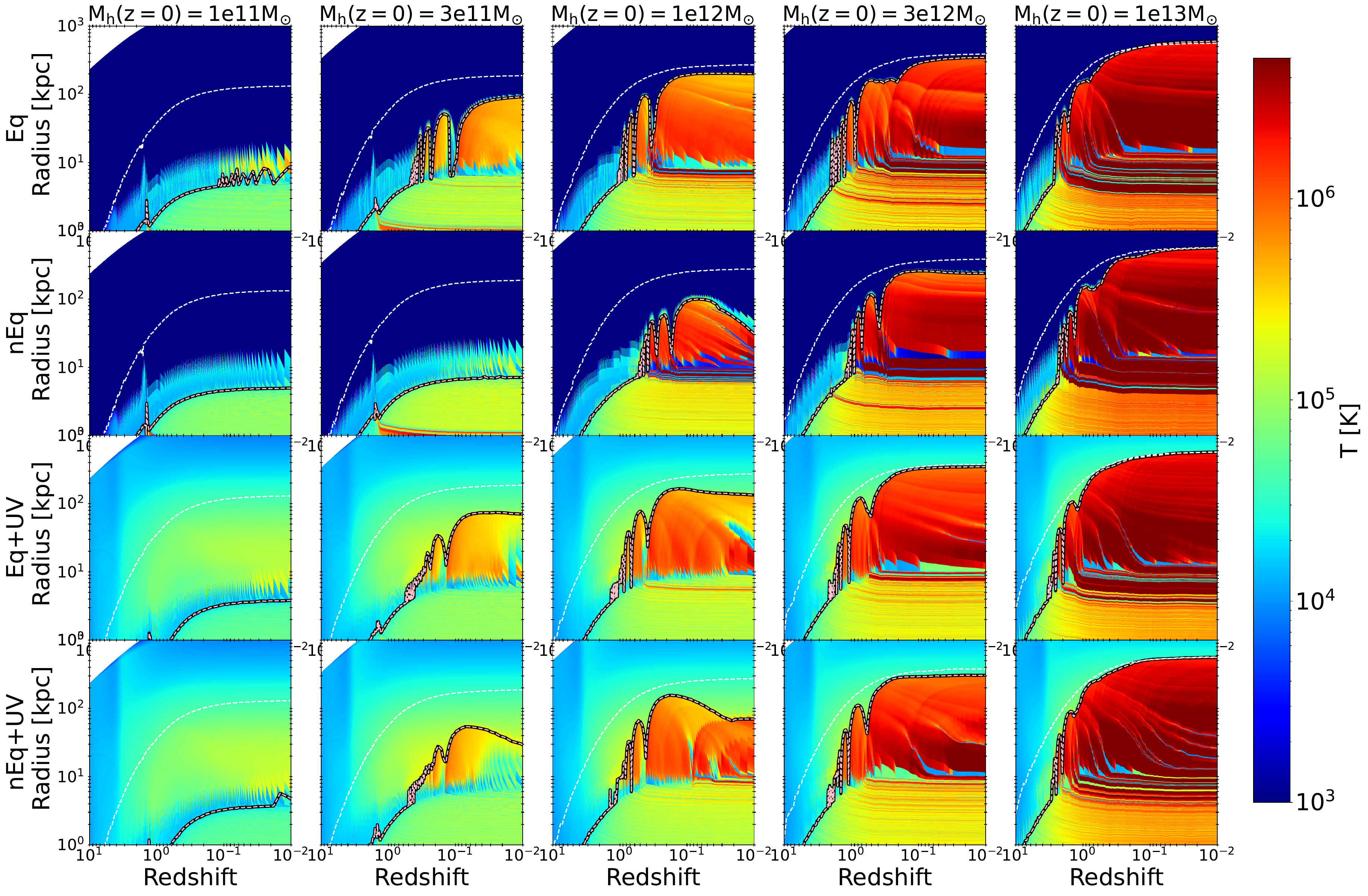}
\caption{Evolution of gas shells in haloes of different masses and physical scenarios. Each row shows a different physical model ('Eq','nEq','Eq+UV','nEq+UV', from top to bottom). Each column is for a different final halo mass ($10^{11}, 3\times10^{11}$, $10^{12}, 3\times10^{12}$, $10^{13}$ \msun from left to right). Each panel shows the trajectories of Lagrangian gas shells in radius–redshift space, colour-coded by gas temperature. The white dashed line indicates the virial radius, and the black-red dashed line indicates the shock radius.}
\label{fig:All_halos_by_T_ZCGM0p3}
\end{figure*}

Figure \ref{fig:All_halos_by_T_ZCGM0p3} shows results from the simulations of all masses and modes. Each panel shows the evolution of a single halo, in terms of its shells' radii at each redshift. Different columns are for different halo masses, and different rows are for the four different physical scenarios. The diagrams are colour-coded by the gas temperature at each radius and redshift. The white dashed line displays the theoretical virial radius inferred from a spherical overdensity criterion\footnote{The virial radius at each time is defined as the radius of the shell whose mean enclosed density exceeds the virial overdensity. For 
a $\Lambda$CDM universe with $\Omega_m=0.3$ and $\Omega_{\Lambda}=0.7$ the virial overdensity at $z=0$ is $\Delta_v\approx340$. See \cite{Bryan_and_Norman_1998}, eq. 6 therein.}.
We use a temperature threshold, $T_{\rm th}=2\times10^5$~K, to continuously identify the shocked region during the simulation. The shock front tracking is shown by the dashed black-red lines in Fig.~\ref{fig:All_halos_by_T_ZCGM0p3}. The identified shock radius closely follows the outer boundary of the hot, post-shock gas in all cases.
The gas infall halts at $\sim0.1,r_{\rm vir}$ due to the prescribed angular momentum. This angular-momentum-supported region corresponds to the stellar and gaseous disc.

%%%%%%%%%%%%%%%%%%%%%%%%%%%%%%%%%%%%%%%%%%%%%%%%%%%%%%%%%%%%%%%%%%%%%%%%%%%%%%%%%%%%%%%%%%%%%%%%%
\subsection{Equilibrium Ionisation (No UV Background, 'Eq')}\label{subsec:equilibrium}

In this scenario, ionisation and radiative cooling are calculated assuming collisional ionisation equilibrium (CIE). This is the same assumption made in \citetalias{Birnboim2003}. 
The evolution of the gas trajectories and temperature for this scenario is shown in the top row of Figure~\ref{fig:All_halos_by_T_ZCGM0p3}, with different panels corresponding to different final halo masses.
With no background radiation, the IGM gas, which initially expands with the Hubble flow and eventually turns around and contracts, remains cold, at $T\sim10$–$100$~K, and evolves only through adiabatic processes.

The temperature only increases when a shock forms. This can be seen in the top row of Fig.~\ref{fig:All_halos_by_T_ZCGM0p3}, where the appearance of the red/orange regions marks the formation of shock-heated gas. In lower-mass haloes these shocks are often unstable and short-lived: radiative cooling removes the post-shock pressure support, causing the shock to recede and the gas to cool again. In the $M_h=3\times10^{11}$~M$_\odot$ case, a shock forms intermittently and remains unstable for much of the evolution, although a more persistent shock appears at later times. By contrast, haloes with $M_h\gtrsim10^{12}$~M$_\odot$ develop stable virial shocks that persist and maintain a hot halo, consistent with the results of \citetalias{Birnboim2003}. Note, that for these high masses, the stable shock radius also lies close to the theoretical virial radius inferred from the spherical overdensity criterion. For the $M_h=10^{11}$~M$_\odot$ halo, no virial shock develops because cooling at the virial radius is too efficient. It is dominated by Ly$\alpha$ emission and strong metal-line cooling, and prevents the shock from forming where it would otherwise be expected.

At high masses ($\gtrsim3\times10^{11}$~M$_\odot$), we occasionally observe transient cooling-flow-like behaviour in the temperature. The top-right panel is an excellent example of such a flow where several baryonic shells cool and clump together to produce a colder and denser set of shells that finally fall down to the galactic disc. Although this picture is not exactly the same as in \citep{Stern2019}, it shows a strong resemblance.

%%%%%%%%%%%%%%%%%%%%%%%%%%%%%%%%%%%%%%%%%%%%%%%%%%%%%%%%%%%%%%%%%%%%%%%%%%%%%%%%%%%%%%%%%%%%%
\subsection{Nonequilibrium Ionisation (No UV Background, 'nEq')} \label{subsec:NEI}

The temperature evolution in the non-equilibrium case is shown in the second row of Fig.~\ref{fig:All_halos_by_T_ZCGM0p3}. As in the equilibrium case, the appearance of red regions indicates the formation of shock-heated gas. Comparing the equilibrium and non-equilibrium cases reveals several important differences for the lower mass haloes. For the $3\times 10^{11}$ \msun halo, no shock forms in the time-dependent case, whereas in the equilibrium case a shock appears at late times. For the $10^{12}$ \msun halo, a shock does form in the non-equilibrium model, but its radius decreases with time and does not approach $R_{\rm vir}$ as it does in the equilibrium case. For haloes with $M\gtrsim 3\times 10^{12}$ \msun, the evolution is qualitatively similar in the two models.
This indicates that in the time-dependent case, the critical mass for stable shock formation is higher than in CIE, with stable shocks appearing only for masses $\gtrsim 10^{12}$ \msun in our simulations.

These differences arise from the time-dependent nature of the ionisation calculation. Following the non-equilibrium ion fractions alters the thermal evolution of the gas because the cooling rates depend on the instantaneous ion populations. In particular, the gas entering the shock is significantly underionised relative to the shock temperature, making cooling more efficient than in the equilibrium case. This is because the lower ionisation species that remain present at the shock temperature, are more efficiently excited by the hot thermal electrons. The enhanced cooling reduces the post-shock pressure support, making the shock more difficult to sustain.

%%%%%%%%%%%%%%%%%%%%%%%%%%%%%%%%%%%%%%%%%%%%%%%%%%%%%%%%%%%%%%%%%%%%%%%%%%%%%%%%%%%%%%
\subsection{Equilibrium Ionisation with UV Background ('Eq+UV')} \label{subsec:EqUV}

The third row in Figure \ref{fig:All_halos_by_T_ZCGM0p3} shows the evolution of haloes in photoionisation equilibrium with the redshift-dependent UV background.
The ionising background significantly alters the thermal and ionisation properties of the IGM gas. Instead of remaining cold and evolving only through adiabatic processes, the gas temperature and ionisation state are now also affected by photoionisation and photoabsorption heating. As a result, the typical IGM temperatures increase to $\sim 10^4-10^5$ K, with the exact equilibrium temperature depending on the gas density and redshift.

We note that in our simulations the gas temperature does not always correspond to the standard thermal equilibrium solution obtained by equating the radiative cooling and photoabsorption heating rates. This is because adiabatic processes associated with cosmic expansion and gravitational collapse also affect the thermal evolution of the gas. In particular, adiabatic cooling due to cosmic expansion often dominates in the low-density IGM, keeping the gas temperature near $10^4$~K. As the gas collapses and approaches the virial radius, adiabatic compression raises its temperature above the standard thermal equilibrium value \citep[e.g.][]{Arya2026}. In both cases, the deviation from equilibrium is at most a factor of a few.

The red and orange regions again mark the formation of shock-heated gas. Despite the substantial differences in the IGM properties between the runs with and without a UV background, the shocked halo properties remain broadly similar, with comparable shock-formation epochs and overall shock stability. This is expected because the properties of the shock-heated gas are dominated by collisional processes at the virial temperature rather than by the background radiation.

%%%%%%%%%%%%%%%%%%%%%%%%%%%%%%%%%%%%%%%%%%%%%%%%%%%%%%%%%%%%%%%%%%%%%%%%%%%%%%%%%%%%%%%%%%%
\subsection{Nonequilibrium Ionisation \& UV Radiation ('nEq+UV')} \label{subsec:neq-uv}

Finally, in the last row of figure \ref{fig:All_halos_by_T_ZCGM0p3}, we display the temperature evolution in haloes exposed to the UV background radiation in the non-equilibrium case.
This scenario is likely the most realistic for astrophysical haloes.

As in the Eq+UV case, the impact of the UV background on the IGM properties is clearly apparent: the temperature evolution of the IGM gas closely matches that of the equilibrium model, since departures from ionisation equilibrium have little effect on the slowly evolving IGM.

When the UV background is included, departures from equilibrium also no longer play a significant role in shaping the structure and stability of the CGM, unlike in the non-equilibrium model without radiation. 
This occurs because the UV background pre-ionises the IGM gas as it approaches the virial shock. As a result, the infalling gas is much less under-ionised than in the radiation-free case, and the associated cooling rates are reduced, approaching their equilibrium values. This allows thermal pressure to build up more steadily behind the shock.

\begin{figure*}
    \centering
    \includegraphics[width=0.98\textwidth]{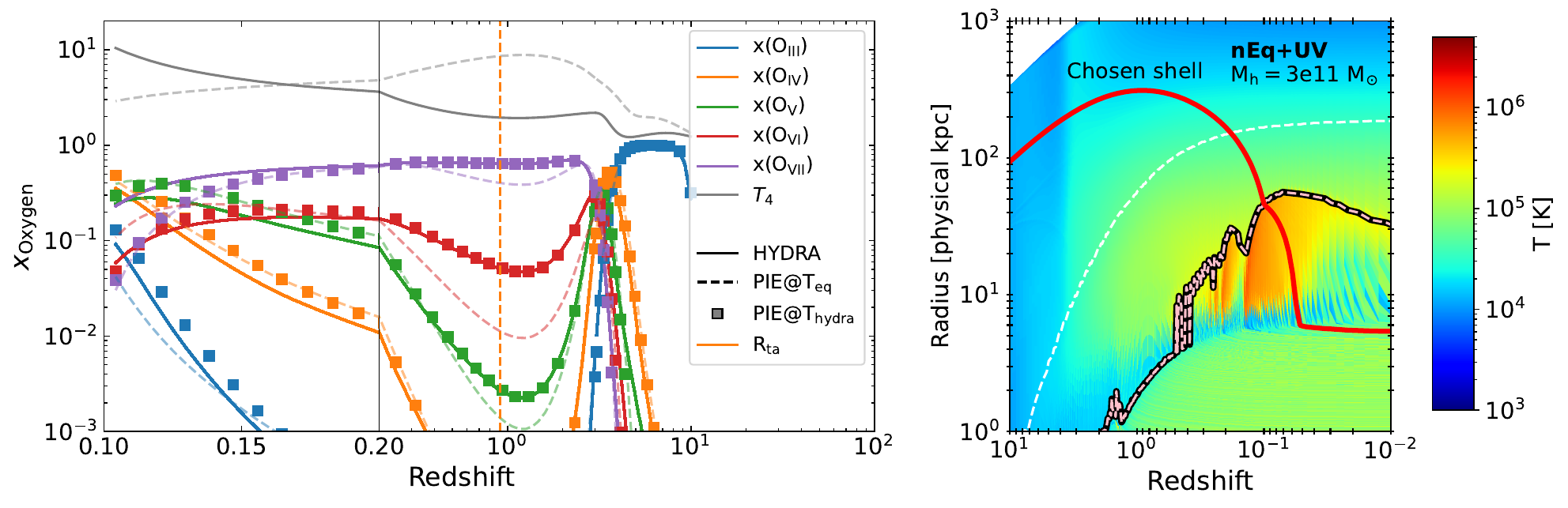}
    \caption{Right: $r-z$ trajectory of a gas shell (no.~441) in a $3\times10^{11}$~M$_\odot$ nEq+UV halo.
Left and middle: Oxygen ion fractions vs. redshift for the same shell (left panel focuses on late times). Gray curves show 
$T_4\equiv T/(10^4 {\rm \,K})$; coloured curves show ion fractions. Solid lines: \hydra; dashed: PIE at $T_{\rm eq}$; squares: PIE at $T_{\hydra}$.}
    \label{fig:x-oxygen}
\end{figure*}

\subsection{Minimal Mass for Shock Formation}

We refer to the lowest halo mass in our grid that develops a persistent, spatially extended shocked region (see Fig.~\ref{fig:All_halos_by_T_ZCGM0p3}) as the critical mass. 
In our simulations, the critical mass is only bracketed by the discrete halo masses considered (spaced by ~0.5 dex), and we therefore quote it to order-of-magnitude accuracy. Across all models, the transition spans approximately one mass interval (~0.5 dex), but its location shifts depending on the ionisation physics.

With this in mind, while non-equilibrium effects alone enhance post-shock cooling and appear to shift the critical mass for shock formation to higher values ($M_h\gtrsim10^{12}$ \msun), the inclusion of the UV background largely suppresses these non-equilibrium effects by pre-ionising the infalling gas. As a result, the critical mass for stable shock formation returns to 
$M_h\sim3\times10^{11}$ \msun, independent of whether the ionisation state is treated in equilibrium or not.
Thus, by pre-ionising the infalling IGM, the UV background moderates the non-equilibrium cooling behind the shock and brings the critical mass back to the equilibrium value found by \citetalias{Birnboim2003}.

\begin{figure*}
\centering
\includegraphics[width=0.98\textwidth]{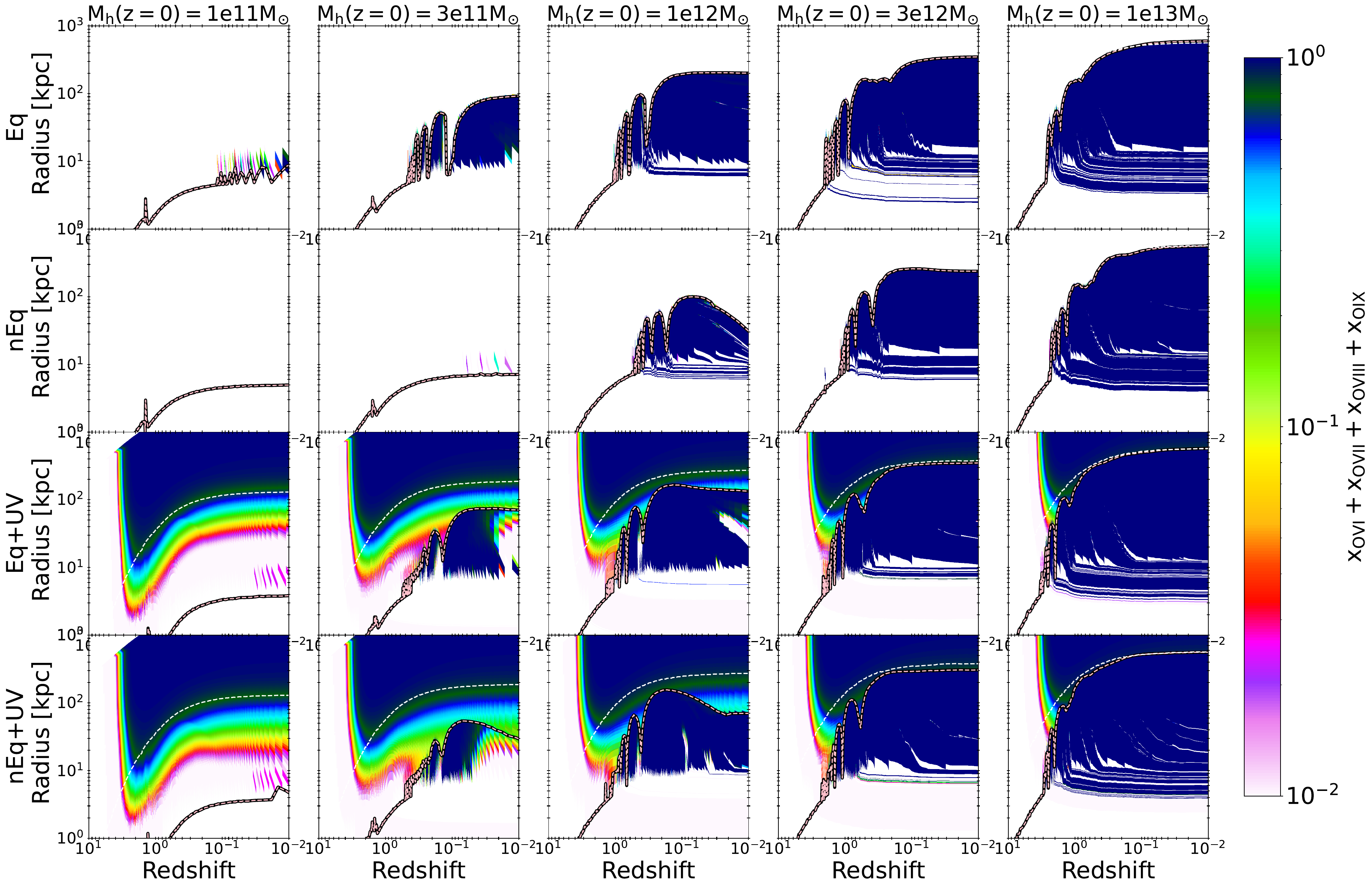}
\caption{Same as Fig.~\ref{fig:All_halos_by_T_ZCGM0p3}, 
but colour-coded by the sum of the ion fractions of \ovi, \ovii, \oviii and \oix. This sum represents the ionisation state of the gas as it approaches the halo.}
\label{fig:All_halos_by_xO_ZCGM0p3}
\end{figure*}

\section{IGM Temperature and Ion Fractions} \label{sec:ionization-state}

In this section, we discuss the thermal state and ionisation structure of the infalling IGM gas in our simulations. We focus on the models that do include the UV background, as we consider those more realistic. 

Two aspects are of particular interest. 
First, we demonstrate that the temperature of the IGM gas 
generally does not follow the standard photoabsorption-heating equals radiative-cooling thermal equilibrium solution. Instead, adiabatic effects associated with cosmic expansion and gravitational compression modify the thermal evolution, producing temperatures that deviate from the usual equilibrium curve.
Second, we show that when the UV background is included, the low-density IGM is a natural site for the production of high ions. As we show in \S~\ref{subsec:ovi}, these IGM ions may significantly contribute to UV absorption around galaxies \citepalias[see also][]{Bromberg_2025}.

In Fig.~\ref{fig:x-oxygen} we show the evolution of the gas temperature (gray) and oxygen ion fractions (coloured curves) as a function of redshift for a representative gas shell as it expands with the universe, turns around, and eventually approaches the virial shock at $z \simeq 0.1$. This is shell no.~441 in the $3\times10^{11}$~\msun nEq+UV model. 
The middle panel shows the evolution starting at $z=10$, while the left panel focuses on the final stages of the shell’s approach to the virial shock. In each panel, the solid curves show the results of our \hydra simulation. The dashed curves show the corresponding results assuming standard thermal and photoionisation equilibrium (PIE): 
Specifically, the dashed gray curve shows the equilibrium temperature ($T_4\equiv T/(10^4~{\rm K})$) obtained by requiring that photoabsorption heating balances radiative cooling at the local gas density for the redshift-dependent radiation field. The dashed coloured curves show the corresponding photoionisation equilibrium ion fractions evaluated at this equilibrium temperature. 
We also plot, with coloured squares, the PIE abundances evaluated at the temperature obtained in our \hydra simulations.

As discussed in \S~\ref{subsec:EqUV}, adiabatic processes associated with cosmic expansion and gravitational collapse affect the thermal evolution of the gas. During the expansion phase, the gas temperature remains lower than the thermal equilibrium value (as seen in the right panel), while during collapse it rises above the equilibrium value (visible in the left-hand panel).
Since the gas temperature departs from the thermal equilibrium value, the associated ion fractions also deviate from the standard equilibrium predictions. In addition, Fig.~\ref{fig:x-oxygen} demonstrates that as the gas approaches the accretion shock, departures from ionisation equilibrium further shift the ion fractions away from the equilibrium values corresponding to the simulated temperature (solid curves vs. squares), but the deviations remain moderate.

A second aspect that we wish to highlight is the prevalence of high ions in and around galactic haloes. We illustrate this by considering the evolution of the combined abundance of high oxygen ions, \ion{O}{6} $+$ \ion{O}{7} $+$ \ion{O}{8} $+$ \ion{O}{9}. Figure~\ref{fig:All_halos_by_xO_ZCGM0p3} shows the evolution of this high-ion abundance in the same radius–redshift parameter space used in Fig.~\ref{fig:All_halos_by_T_ZCGM0p3}. As expected, high ions are abundant inside the shock-heated CGM for all halo masses and physical scenarios due to efficient collisional ionisation. In the IGM, however, the different models show markedly different behaviour. When no UV radiation is included, the IGM remains cold and largely devoid of high ions, so any contribution to high-ion absorption around galaxies arises from the CGM. By contrast, when the UV background is included, photoionisation drives the ionisation state of the IGM to much higher values, producing significant abundances of high ions outside $R_{\rm vir}$, with substantial contributions between $R_{\rm vir}$ and $R_{\rm shock}$ for haloes with $M_h\gtrsim3\times10^{11}$~\msun. We do not see much difference in the overall ionisation state between the Eq+UV and nEq+UV cases. This pre-ionisation reduces the non-equilibrium cooling enhancement behind the shocks.  This gives rise to IGM contributions to UV absorption around galaxies \citepalias[see][and \S~\ref{sec:columns}]{Bromberg_2025}.

%%%%%%%%%%%%%%%%%%%%%%%%%%%%%%%%%%%%%%%%%%%%%%%%%%%%%

\section{Results: Observational Signatures} \label{sec:columns}

In this section we study the absorption-line signatures that arise in our spherical halo models. As in the halo stability analysis presented above, we now examine how departures from ionisation equilibrium and the presence of a UV background affect the metal absorption properties of galactic environments. Various factors, including the position and stability of the virial shock, as well as the evolving thermal and ionisation state of the gas, affect the observable metal-ion column densities and the relative contributions of the CGM and the IGM to each absorption line.
Here we compute the column densities produced in our spherical models for the different physical scenarios considered in this work. In \S~\ref{subsec:ovi} we examine \ion{O}{6} columns, a commonly used tracer of CGM gas. 
In \S~\ref{subsec:Lya} we focus on the hydrogen Ly$\alpha$ column densities, a primary indicator of IGM absorption systems. Finally, in \S~\ref{subsec:CIV} we consider \ion{C}{4} absorption.

\subsection{\ovi Column Densities} \label{subsec:ovi}

\begin{figure*}
\centering
\includegraphics[width=0.95\textwidth, clip=True, trim={0 0cm 0 0.85cm}]{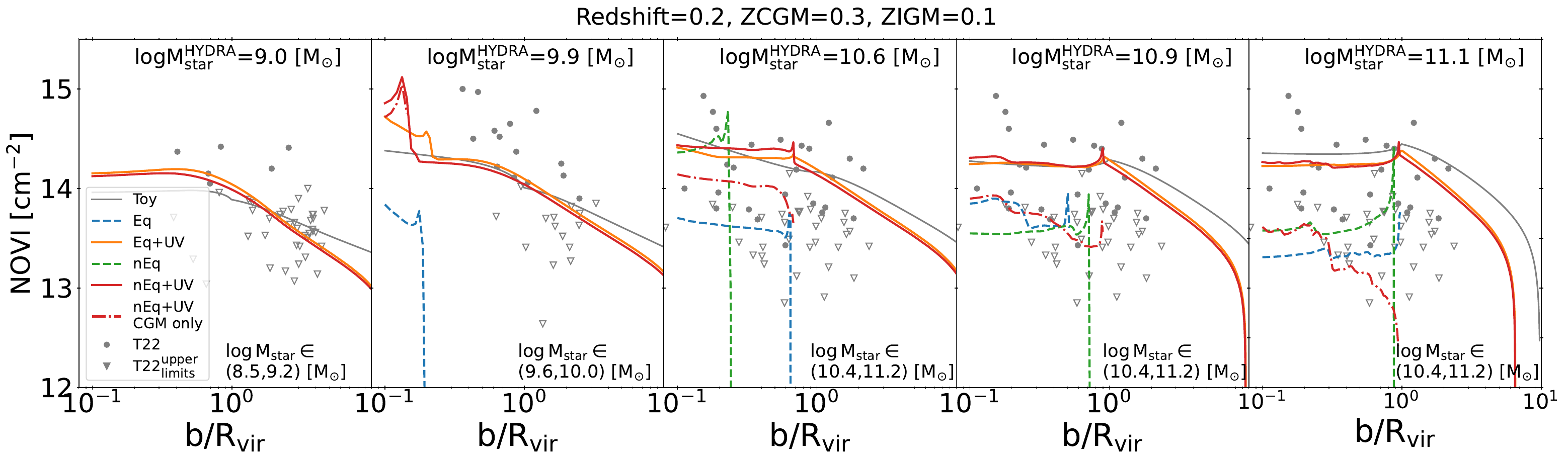}
\includegraphics[width=0.98\textwidth, clip=True, trim={0 0cm 0 0.85cm}]{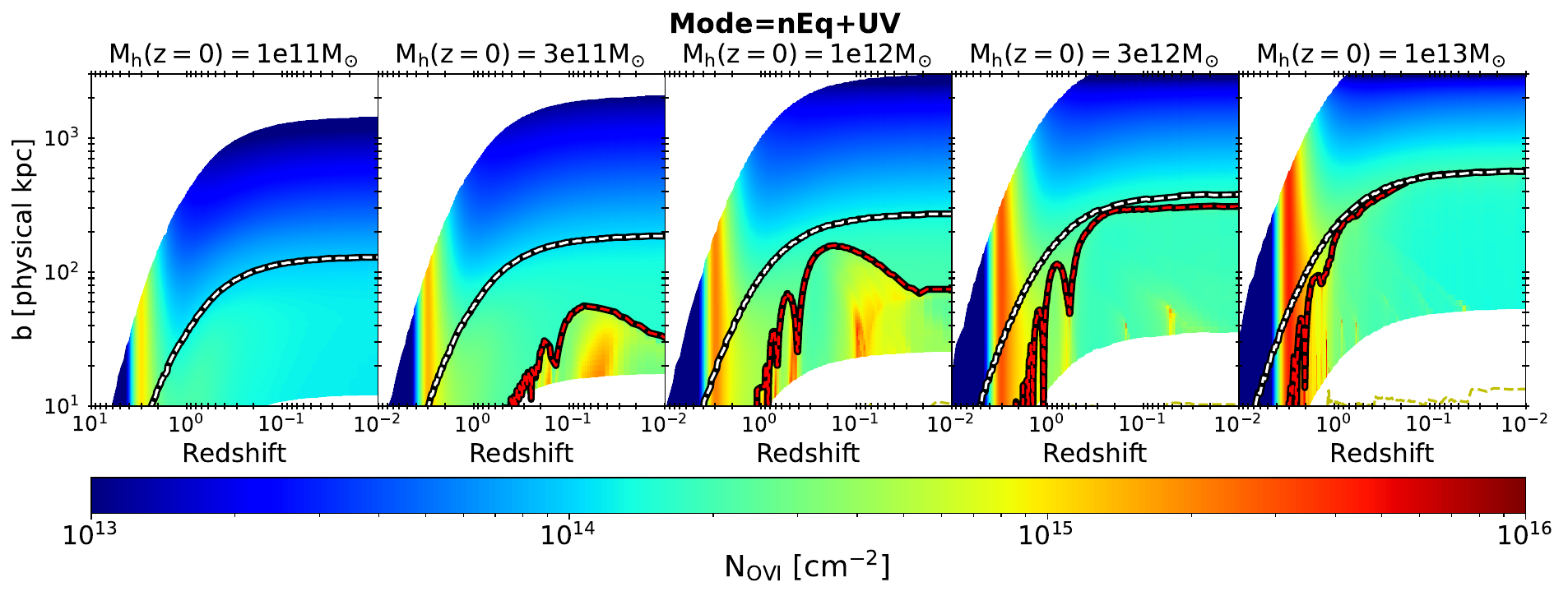}
\caption{
\ion{O}{6} columns. 
\textbf{Top row}: \ion{O}{6} columns at $z=0.2$ as a function of scaled impact parameter, $b/R_{\rm vir}$, for different halo masses (panels). Curves show the models: Eq (blue), Eq+UV (orange), nEq (green), nEq+UV (solid red), the CGM-only contribution ($r<r_s$) to the nEq+UV case (dash–dotted red), and the toy model of \citetalias{Bromberg_2025} (gray). Markers show observational detections (filled circles) and upper limits (empty triangles) from \citet{Tchernyshyov_2022}.
\textbf{Bottom row}: \ion{O}{6} columns for the nEq+UV model across the full $r-z$ parameter space. Colour indicates the column density for each redshift–sightline (impact parameter) combination (see colour-bar).
}
\label{fig:OVIvs_b}
\end{figure*}

In Fig.~\ref{fig:OVIvs_b} we present the \ion{O}{6} column densities obtained from our simulations. In the top row we show a snapshot at 
$z=0.2$ of the integrated \ion{O}{6} column density as a function of the scaled impact parameter, $b/R_{\rm vir}$. Each panel corresponds to a different halo mass. The corresponding stellar masses, derived using the stellar-to-halo mass relation of \cite{Girelli2020}, are indicated within each panel.
As in \citetalias{Bromberg_2025}, we truncate the integrated column densities at a radius of 3 Mpc. 
Adopting a common truncation radius well outside $R_{\rm vir}$ allows a consistent comparison between haloes of different masses.

The different curves represent the physical scenarios considered in this work: Eq (dashed blue), Eq+UV (solid orange), nEq (dashed green), and nEq+UV (solid red). For the nEq+UV case, we also show - with red dash-dotted lines - the column densities arising from the CGM alone (i.e., integrating the column only out to $r_s$). For comparison, the solid gray curves show the column densities predicted by the simplified toy model of \citetalias{Bromberg_2025}. Observational detections (filled circles) and upper limits (empty triangles) from the \citet{Tchernyshyov_2022} compilation of CGM surveys are also plotted.

Fig.~\ref{fig:OVIvs_b} confirms that the inclusion of a UV background alters the state of the infalling IGM. In simulations without a UV background, the IGM remains cold and largely neutral, resulting in very low abundances of highly ionised species. When a UV background is included, however, the IGM becomes warm and significantly ionised, producing substantial populations of high ions even outside the virialised halo.
This difference is clearly reflected in the predicted \ion{O}{6} column densities. Models that include the UV background (red, orange) reproduce the observed columns reasonably well, whereas simulations without an ionising background (blue, green) show a sharp drop in column density at $r_s$, and the columns at larger impact parameters significantly underpredict the observations. The metagalactic radiation field therefore creates an extended, photoionised IGM surrounding the halo that contributes substantially to the absorption along the line of sight.

The top panels of Fig.~\ref{fig:OVIvs_b} also show the \ion{O}{6} column densities produced solely within the CGM for the nEq+UV model (dash–dotted red curves). For the higher-mass haloes (the three panels on the right), the CGM contribution alone does not reproduce the observed \ion{O}{6} columns even within $r_s$, falling short by roughly an order of magnitude compared to the data. Our one-dimensional $\Lambda$CDM hydrodynamical simulations therefore confirm the conclusions of \citetalias{Bromberg_2025}, where a simplified toy model was used to explore the contribution of the surrounding IGM to the observed warm-ion absorption around galaxies. In both cases, we find that the IGM alone can produce column densities comparable to observations.
The results of \citetalias{Bromberg_2025} are shown for comparison by the gray curves\footnote{Note that the metallicity adopted by \citetalias{Bromberg_2025} was three times lower than in our models, compensating for the higher densities produced by their matter-dominated Einstein–de Sitter cosmology, which contains roughly three times more matter.}.

Also apparent in the figure is that the \ion{O}{6} column densities in the Eq+UV and nEq+UV models are nearly identical across all halo masses, both within and outside the shock radius ($r_s$, where the CGM-only column sharply drops).
This similarity arises because, at large impact parameters ($b/R_{\rm vir}>1$), the column densities are dominated by the infalling IGM, where the ionisation state is set primarily by the UV background. Since both models are exposed to the same radiation field, and the ionisation and recombination timescales are much shorter than the (cosmological) timescales for changes in the gas temperature or radiation field, the gas remains close to ionisation equilibrium. 
As a result, departures from equilibrium have little effect on the predicted ion fractions in the IGM.
This is consistent with Fig.~\ref{fig:x-oxygen}, which shows similar \ion{O}{6} abundances in the nEq (solid) and equilibrium (squares, dashed curves) cases. 

Crossing the shock front could, in principle, be a site of significant departures from ionisation equilibrium, since the infalling IGM may be underionised with respect to the shock temperature. However, as shown in Fig.~\ref{fig:All_halos_by_T_ZCGM0p3}, the UV background pre-ionises the IGM, and the warm photoionised gas does not undergo significant cooling after passing through the shock. Further downstream, departures from equilibrium could arise if the shocked gas cools on a timescale shorter than the recombination time. At $z=0.2$, however, Fig.~\ref{fig:All_halos_by_T_ZCGM0p3} shows no substantial cooling behind the shock. We therefore do not expect large differences between the Eq+UV and nEq+UV cases. And indeed, Fig.~\ref{fig:OVIvs_b} confirms that the predicted \ion{O}{6} column-density profiles are nearly identical in the two models, even inside $r_s$.

\begin{figure*}
\centering
    \includegraphics[width=0.95\textwidth]{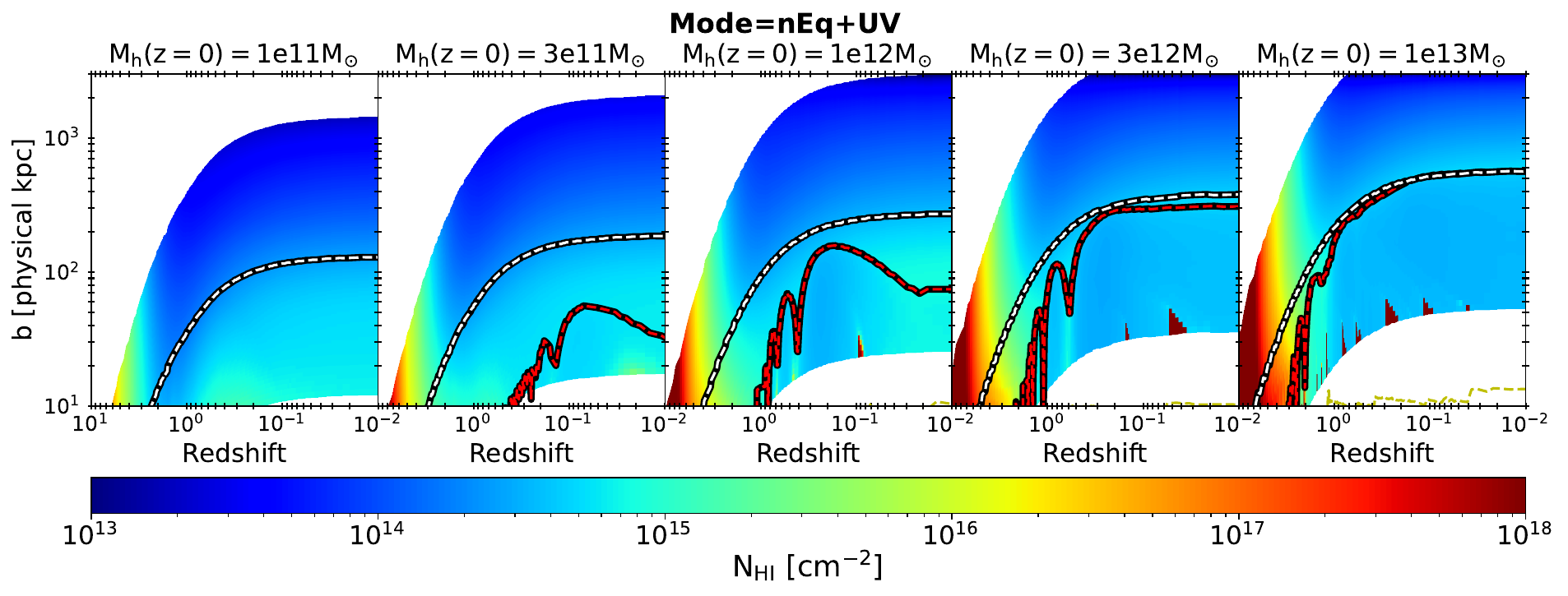} 
    \includegraphics[width=0.95\textwidth, clip=True, trim={0 0cm 0 0.8cm}]{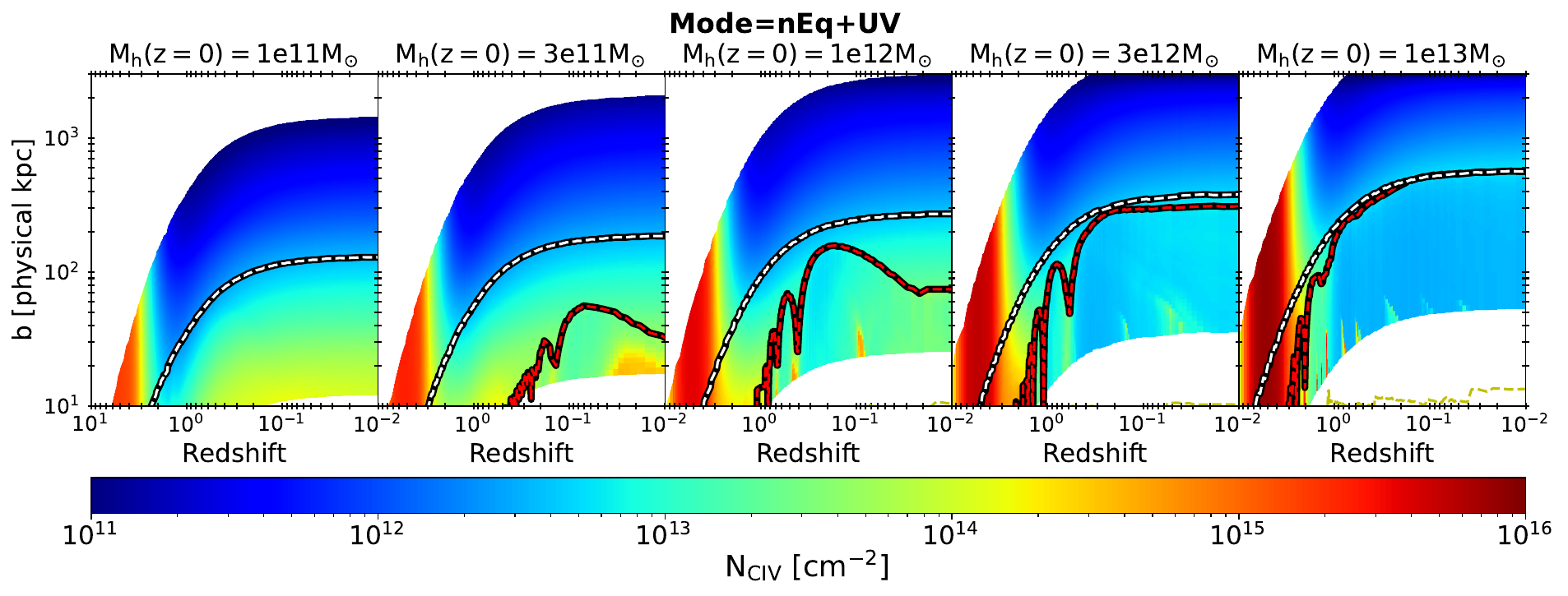}
    \caption{N$_{\rm H\,I}$ (top) and N$_{\rm C\,IV}$ (bottom) columns for the nEq+UV model across the full $r-z$ parameter space. Colour indicates the column density for each redshift–sightline (impact parameter) combination (see colourbar).
    } 
    \label{fig:metal-columns}
\end{figure*}

The bottom row in Fig.~\ref{fig:OVIvs_b} shows the predicted \ion{O}{6} column densities across the full $r-z$ parameter space of Fig.~\ref{fig:All_halos_by_T_ZCGM0p3} for the nEq+UV case. As before, different panels are for different masses, and the virial and shock radii are indicated in each panel. Here the colour scheme shows the \ion{O}{6} column density corresponding to each $r-z$ combination, where the radial coordinate represents the projected sightline through the halo, i.e., the observational impact parameter, $b$.

Fig.~\ref{fig:OVIvs_b} shows that readily observable \ion{O}{6} column densities, $N_{\rm O\,VI}\gtrsim10^{13}$~cm$^{-2}$, are ubiquitously produced in our simulations, extending to several times the virial radius. The overall \ion{O}{6} column density varies only weakly with halo mass, reflecting the dominant contribution of the IGM.
The \ion{O}{6} columns become particularly large at $z=2-4$, reaching $N_{\rm O VI}\gtrsim10^{15}$~cm$^{-2}$, during the peak of cosmic star formation (when the UV background is maximal), as indicated by the orange–red vertical strip in the figure. Such column densities are well above the typical detection limits of optical surveys probing the high-redshift IGM \citep[e.g.][]{Simcoe2002,Simcoe2004,Carswell2002,Bergeron2002,Schaye2000}, and are consistent with observations of highly ionised gas at $z\gtrsim2$. The abrupt transition from low ionization to high ionisation oxygen corresponds to the reionisation of the universe, akin to the well-known hydrogen reionisation.

Finally, we note that while the \ion{O}{6} column densities are generally not strongly affected by departures from ionisation equilibrium, this is not always the case. As discussed above, when the UV background pre-ionises the infalling IGM, departures from equilibrium near the shock radius remain small. In most cases, the shocked gas remains hot, preventing significant non-equilibrium effects associated with rapid cooling. This is the case for $z=0.2$, displayed in the upper row of Fig.~\ref{fig:OVIvs_b}. 
However, Fig.~\ref{fig:All_halos_by_T_ZCGM0p3} shows that rapid cooling can occasionally develop in the post-shock layers for the non-equilibrium models. For example, a large cool region is visible in the $3\times10^{11}$~\msun halo at $z<0.1$, and a localised cooling flow appears in the $10^{12}$~\msun halo at $z\sim0.08$. In such regions, the cooling time may become shorter than the recombination time, leading to departures from ionisation equilibrium \citep[e.g.][]{Gnat2007,Gnat2009}. 
These “flows” arise from instabilities, and their detailed properties are likely numerical in origin. We defer a more robust analysis of these features to future work.

\begin{figure}
    \centering
    \includegraphics[width=1\linewidth]{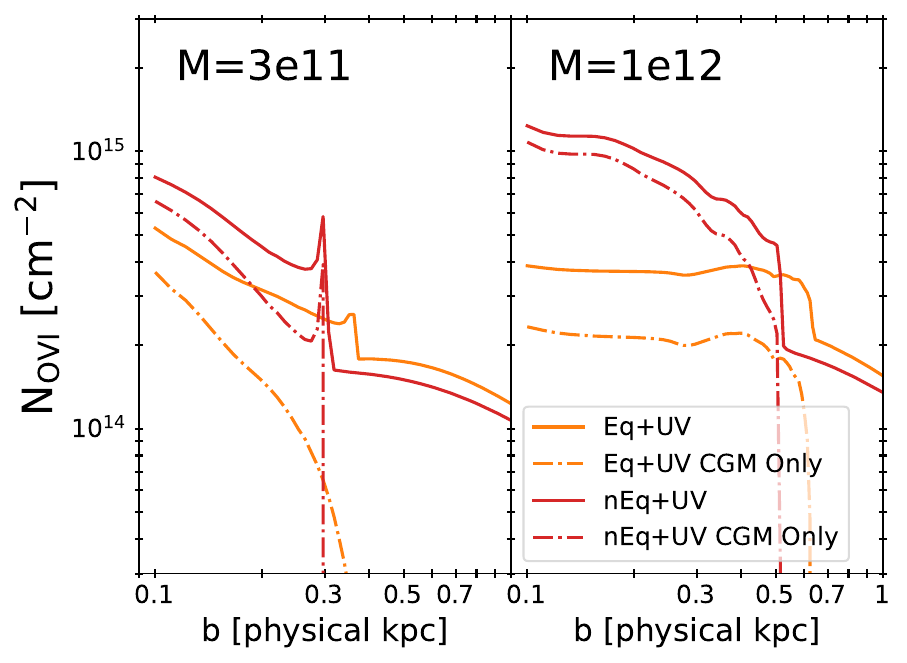}
    \caption{$N_{\rm O\,VI}$ versus impact parameter, as in Fig.~\ref{fig:OVIvs_b}, for $3\times10^{11}$ and $10^{12}$~M$_\odot$ haloes at $z=0.08$, where transient “cooling flow” behaviour is observed.}
    \label{fig:cooling_flow}
\end{figure}

Under such "cooling-flow" conditions, recombination may lag behind cooling, enhancing the columns of high ions. Such enhancements are visible in the lower panel of Fig.~\ref{fig:OVIvs_b} for the relevant halo masses and redshifts. In Fig.~\ref{fig:cooling_flow} we present the \ion{O}{6} column density as a function of impact parameter for the relevant halo masses ($3\times10^{11}$ and $10^{12}$~M$_\odot$) at $z=0.08$, comparing the Eq+UV (orange) and nEq+UV (red) models. As before, solid curves show the total integrated column density, while dash–dotted curves indicate the CGM-only contribution. The figure demonstrates that the CGM columns differ between the equilibrium and non-equilibrium models, producing differences of a factor of $2-3$ in the total observed columns.

\subsection{\hi Column Densities}\label{subsec:Lya}

Ly$\alpha$ absorption is one of the primary tracers of absorber redshift in the IGM. It is particularly useful at high redshifts, where metal lines are less common due to the low metallicity of the early Universe. Figure~\ref{fig:metal-columns} (top panel) shows the \hi\ column density maps from our simulations.
At $z\sim3$, our predicted column densities, $N_{\rm HI} \sim 10^{15-17}$~\psqcm, are consistent with observational constraints on \ion{H}{1} in the IGM \citep[][$N_{\rm HI} \gtrsim 10^{14.5}$~\psqcm]{Songaila1996}. At lower redshifts, $z\sim1$, our models predict $N_{\rm HI} \lesssim 10^{14.3}$~\psqcm in the nearby IGM ($b/R_{\rm vir} \lesssim 3$).

In the CGM, it is generally understood that neutral hydrogen originates in colder “pockets” within a multiphase medium, which are not captured in our simplified spherical simulations. Our models include only the warm–hot phase, and therefore produce relatively small amounts of \hi, with little variation in $N_{\rm HI}$ across the CGM. While the resulting \ion{H}{1} column densities are consistent with the lower end of detections in the COS-HALOS survey \citep[][$N_{\rm HI} \sim 10^{14-19}$~\psqcm]{Prochaska2017}, they do not reproduce the higher observed values. This is expected, and supports the standard picture in which neutral hydrogen arises from a multiphase CGM.

\subsection{\civ Column Densities}\label{subsec:CIV}

The \civ $\rm 1548\AA$ transition is a commonly observed absorption line in quasar spectra at $z\sim 2-3$, and is widely used to constrain the metallicity and ionisation state of the IGM. Observations typically find \civ column densities of $N_{\rm C\,IV} \sim 10^{12}-10^{14}$ \psqcm \citep{Songaila1996, Schaye2000, Simcoe2011, Banerjee2023}, often associated with partial Lyman limit systems (pLLSs; $N_{\rm HI} \sim 10^{16}-10^{17}$ \psqcm) and Lyman Limit Systems (LLS; $N_{\rm HI} \sim 10^{17}-10^{19}$ \psqcm).

Our simulations reproduce such systems, as shown in the bottom panel of Fig.~\ref{fig:metal-columns}. At $z\sim 2-3$, \civ is abundant ($N_{\rm C\,IV}\sim 10^{13}-10^{15}$~\psqcm) in gas with $N_{\rm HI}\sim 10^{15}-10^{17}$~\psqcm, consistent with observational constraints. In addition, \civ is present over a wide radial range, extending well beyond the virial radius and out to $\sim10 R_{\rm vir}$, in agreement with the findings of \citet{Banerjee2023}.

At lower redshifts ($z\lesssim1$), we find \civ columns of $N_{\rm C\,IV}\sim 10^{13}-10^{14}$ \psqcm around low-mass haloes , primarily arising outside the shock radius.
For low-mass haloes , virial shocks are often unstable, and when they do form they typically lie well within $R_{\rm vir}$. As a result, even sightlines with small impact parameters ($b<R_{\rm vir}$) do not probe the virialised CGM, but instead intersect the infalling IGM.
For higher-mass haloes , the \ion{C}{4} column densities decrease by roughly an order of magnitude, but may still be detectable in absorption against background AGN.

%%%%%%%%%%%%%%%%%%%%%%%%%%%%%%%%%%%%%%%%%%%%%%%%%%%%%%%%%%
\section{Caveats} \label{sec:discussion}

The hydrodynamical simulations presented in this work represent a significant step forward over the toy-model predictions of \citetalias{Bromberg_2025}. In particular, the underlying cosmological framework is the $\Lambda$CDM model, which accounts for both the reduced matter density ($\Omega_m=0.3$) and the accelerated expansion driven by the cosmological constant. In addition, the hydrodynamical evolution of the gas is treated self-consistently, allowing us to follow shock formation and stability while properly capturing the associated jump conditions. Finally, departures from ionisation equilibrium and the corresponding non-equilibrium cooling are explicitly included.

Despite these improvements, our one-dimensional spherical simulations do not capture many of the complexities of real astrophysical haloes. In particular, departures from spherical symmetry, as well as internal sources that drive winds and inject energy and radiation, are not included in our models. We discuss these limitations below.

\subsection{Effects of Non-Spherical Dynamics}
While our current one-dimensional framework provides a useful baseline for understanding halo thermodynamics, it is important to delineate the limitations imposed by the assumption of smooth, isotropic accretion. In the standard 
$\Lambda$CDM paradigm, dark matter haloes reside in highly anisotropic large-scale environments. The dynamical and radiative processes that govern galaxy formation are  inherently non-spherical and often unstable, and lead to interactions between different thermal phases, producing a multiphase medium. Cold accretion filaments penetrate hot haloes , energetic outflows may be launched from galactic discs, and satellite systems continuously interact with the surrounding gas. Such "local" processes play a key role in structure formation and significantly disturb the gas dynamics and composition. As a result, realistic galactic haloes are complex systems that can differ substantially from our idealised spherical models \citep[see, e.g.,][]{Oppenheimer2018}.
This work is not intended to model the full multiphase CGM, but rather to isolate how ionisation physics alone affects shock stability and large-scale ion distributions.

\subsection{Effects of Supernovae/AGN Feedback on Virial Shocks}
Our simulations neglect the extensive impact of feedback processes. Central galactic outflows driven by star formation bursts or Active Galactic Nuclei (AGN) inject significant amounts of energy, momentum, and metal-enriched material into the CGM \citep{Springel2003,Fabian2012}. These feedback mechanisms can fundamentally alter the cooling function, $\Lambda(T,Z)$, by redistributing metals within and beyond the virial radius ($R_{\text{vir}}$) and by suppressing catastrophic cooling flows \citep{Schawinski2007,Sutherland1993,Tumlinson2017}.

The interplay between filamentary inflows and feedback-driven outflows can also generate thermal instabilities and drive departures from ionisation equilibrium, potentially modifying both the shock stability and absorption-line signatures \citep{Fielding2017}. Additionally, radiation generated by star-forming regions can also suppress radiative cooling behind the virial shock \citep{Sarkar2022}. 
Taken together, these effects suggest that both the stability of the virial shock and the resulting absorption-line signatures may differ from those predicted by our idealised model. A more complete treatment therefore requires further investigation using more comprehensive hydrodynamic simulations and surveys. Incorporating energy and radiation feedback on the CGM physics will be the focus of future work. 
While this limitation is important, we note that the current state-of-the-art 3D cosmological simulations vary wildly in their subgrid and feedback prescriptions.
This leads to significantly different predictions, except for those features that the models have been explicitly tuned to reproduce.
We therefore consider this limitation to be shared by most galaxy formation simulations.

\subsection{Sensitivity to the UV Background and Metallicity}

Our calculations adopt the \citet{Haardt2012} metagalactic UV background. Alternative UV-background synthesis models, such as \cite{Puchwein2019} and \cite{FG20}, predict somewhat different photoionisation and photoabsorption heating rates, typically differing by factors of a few ($\lesssim2–3$). We also adopt a simplified metallicity prescription, with fixed values for the IGM and CGM. These values are observationally motivated but are not unique, and represent a broader range of plausible metallicities.

Both the UV background model and the adopted metallicity affect the thermal state and ionisation balance of the gas. In the context of shock stability, varying either is therefore expected to shift the quantitative value of the transition mass for stable shock formation in all four computational modes. However, we expect the qualitative result to remain unchanged: non-equilibrium ionisation in initially cold gas enhances post-shock cooling and reduces post-shock pressure support, making stable shocks more difficult to sustain, whereas pre-ionisation by a metagalactic UV background mitigates this effect and restores shock stability.

The predicted metal-line column densities are also sensitive to these assumptions. In particular, in \citetalias{Bromberg_2025} we demonstrated that metal-ion columns scale approximately linearly with metallicity in both the CGM and IGM. These uncertainties therefore primarily affect normalization and threshold values, rather than the underlying physical mechanism identified here.

\section{Conclusions}
In this paper, we build upon the one-dimensional, spherically symmetric Lagrangian hydrodynamics code \hydra (\citetalias{Birnboim2003}, \citealp{Dekel2006,Birnboim2007}) and extend it to include the effects of the redshift-dependent metagalactic UV background and departures from ionisation equilibrium \citepalias{Sarkar2021a} on the thermal evolution, stability, and absorption-line signatures of galactic haloes.

To isolate the impact of these additional physical ingredients, we compare the evolution of several models. As a baseline, we consider an equilibrium model without a UV background (Eq), similar to that presented in \citetalias{Birnboim2003}. We then introduce each effect separately: a radiation-free model with non-equilibrium ionisation and cooling (nEq), and an equilibrium model that includes the UV background (Eq+UV). Finally, we combine both effects in a single model (nEq+UV), which provides the most realistic representation of astrophysical haloes.

We summarise our main results and conclusions as follows:

\begin{itemize}
\setlength{\itemsep}{0pt}
\setlength{\topsep}{0pt}
\setlength{\parsep}{0pt}
\setlength{\partopsep}{0pt}
\setlength{\leftmargin}{1em}
    \item Using the equilibrium radiation-free model, we reproduce the results of \citetalias{Birnboim2003}, and recover the existence of a critical halo mass $\sim3\times10^{11}$~M$_\odot$, below which stable shocks never form.
    \item Non-equilibrium ionisation alone (nEq) reduces virial shock stability. This is because as the cold, neutral infalling IGM is shocked, it is significantly underionised relative to the post-shock temperature, and the resulting enhanced cooling reduces the post-shock pressure support. This shifts the critical mass for stable shock formation to higher values $\sim10^{12}$~M$_\odot$.
    \item The metagalactic UV background pre-ionises and heats the infalling IGM, producing an extended layer of warm, ionised gas extending well beyond the virial radius.
    \item We find that departures from equilibrium are largely suppressed once the UV background is included. When the pre-ionised IGM is shocked, the non-equilibrium cooling enhancement largely disappears, returning the critical mass to the \citetalias{Birnboim2003} value, $\sim3\times10^{11}$~M$_\odot$.
    \item We find that the temperatures in the infalling IGM are not given by the standard photoabsorption heating–cooling thermal equilibrium solution, but are instead modified by cosmic expansion and gravitational collapse.
    \item We show that our simulations produce \ovi column densities comparable to observations in and around galactic haloes. We confirm the results of \citetalias{Bromberg_2025}, that (in spherical models) the IGM alone can produce \ion{O}{6} absorption comparable to observations. Furthermore, the \ion{O}{6} column densities depend only weakly on halo mass.
    \item Non-equilibrium ionisation effects are generally small in the presence of a UV background, and departures from equilibrium typically do not strongly affect the observational signatures, except in rapidly cooling transient regions. 
    \item We also find that at $z\sim 3$, the photoionised IGM around galaxies can give rise to \hi and \civ columns of $\sim10^{15-17}$~\psqcm and $10^{13-15}$~cm$^{-2}$, respectively, consistent with several blind absorption-line surveys.
\end{itemize}

Although our approach is highly idealised compared to full three-dimensional cosmological simulations, our framework allows us to isolate and control the underlying microphysics, providing a clearer understanding of the processes governing metal-line absorption around galaxies. By systematically incorporating additional physical ingredients, we assess the role of each component in shaping the thermal and ionisation structure of the gas. Future extensions of this work - including feedback, metal transport, and local radiation fields - will enable a more complete and controlled exploration of these effects, complementing the statistical insights from large cosmological simulations.

% %%%%%%%%%%%%%%%% MAIN BODY ENDS HERE %%%%%%%%%%%%%%%%
\section*{Acknowledgements}
This research was supported by the ISRAEL SCIENCE FOUNDATION (ISF grant No. 2527/25).

\section*{Data Availability}
The data underlying this article will be shared on reasonable request to the corresponding author.

\vspace{2cm}
\bibliographystyle{mnras}
\bibliography{merged_bibliography_nodup}

\label{lastpage}
\end{document}